# Optoelectronic switching of addressable molecular crossbar junctions

J. C. Li*

*Department of Chemistry, The University of Chicago, Chicago, IL 60637*

(2006-Nov-22)

This letter reports on the observation of optoelectronic switching in addressable molecular crossbar junctions fabricated using polymer stamp-printing method. The active medium in the junction is a molecular self-assembled monolayer softly sandwiched between gold electrodes. The molecular junctions are investigated through current-voltage measurements at varied temperature (from 95 to 300 K) in high vacuum condition. The junctions show reversible optoelectronic switching with the highest on/off ratio of 3 orders of magnitude at 95 K. The switching behavior is independent of both optical wavelength and molecular structure, while it strongly depends on the temperature. Initial analysis indicates that the distinct binding nature of the molecule/electrode interfaces play a dominant role in the switching performance.

**Corresponding author. E-mail: jianchang_li@hotmail.com

Metal/molecule/metal junction is one of the most useful test-beds in studying the charge transport properties of molecular electronic devices in which organic molecules act as the functional units. So far, molecular electronic components such as switches, rectifiers and gates have been widely investigated using a variety of methods.[1-8] However, the relationship between the device performance and molecule-specific properties is still not well understood partially owing to poorly defined metal/molecule interfaces. This is quite true for molecular junctions where a metal electrode has to be thermally evaporated onto the target organic monolayer. Although the approaches of mercury droplet and scanning probe microscope tip based molecular junctions were shown to be simple and non-destructive methods for studying organic monolayers,[9-11] they are far from practical applications.[12] A molecular junction can be used as a practical device only if it is addressable. It is desirable to develop an effective way to fabricate addressable molecular junctions. On the other hand, despite a number of experimental and theoretical advances in understanding the charge transport of molecular systems, major challenges still remain.[13,14] One big issue is how to optically control a molecular junction. Can we realize laser controlled molecular switches and transistors?[15] Previous studies were mostly focused on DC current–voltage ($I$–$V$) characteristics of molecular systems, AC transport properties are not examined in detail yet.

Here, a polymer stamp-printing method has been developed to fabricate addressable molecular crossbar junctions, which provides a simple way to obtain optoelectronic switches and transistors driven by a low frequency AC electrical field. Figure 1(a) shows how the junctions were fabricated. First, an Au electrode pattern (40 nm-thick) was formed on a 1 cm × 1 cm undoped Si (100) substrate using photolithography. The target molecule was then self-assembled onto the Au surface. An elastomeric poly(dimethylsiloxane) (PDMS) stamp was used to print the top electrodes.[16] The stamp was prepared by casting and curing prepolymer against a negative photoresist mold.[17] A 20 nm-thick Au film was deposited onto the stamp (cooled by liquid nitrogen) at pressure of $1\times10^{-6}$ Torr and rate of 1 nm/s. Under microscope, the stamp was brought into contact with the pre-fabricated substrate/Au bars with self-assembled monolayers (SAMs). The substrate/stamp set was placed into vacuum and cooled to 95 K for 10 minutes. After warming up, the stamp was carefully removed and the Au electrodes were transferred onto the substrate with an overall good junction rate of 15 per cent. Figure 1(b) shows several as-fabricated crossbar junctions. Finally, the sample was loaded back into the vacuum chamber and low frequency (0.002 Hz) AC $I$–$V$ characteristics were measured using a HP 3325A Synthesizer/Function generator and a Keithley 619 electrometer.

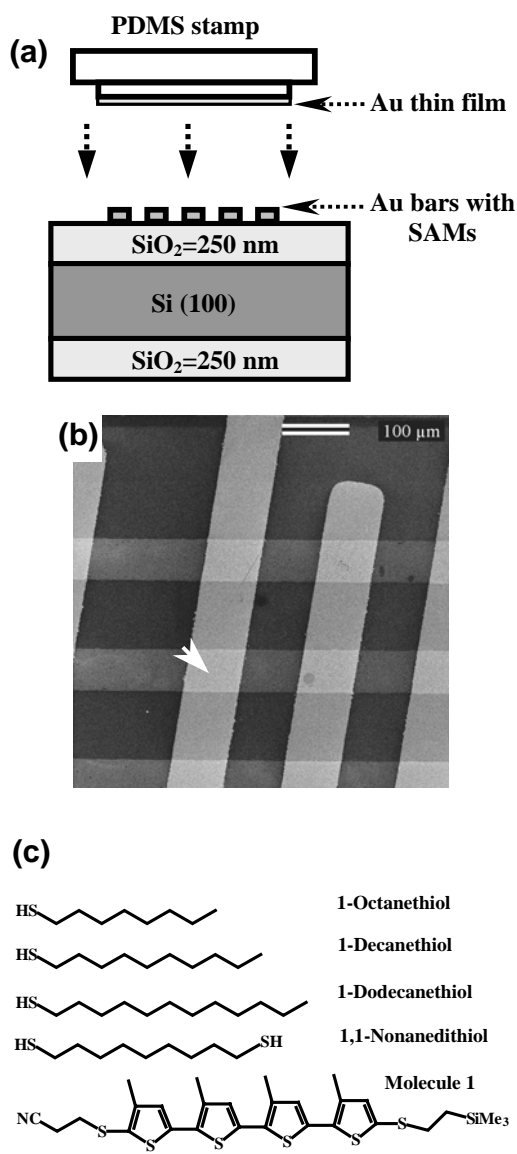

**Figure 1** (a) Schematic drawing shows how the molecular crossbar junctions are stamp printed. (b) Scanning electron microscopy image of several crossbar junctions (with one pointed by the arrow). (c) Chemical structure of the molecules studied.

Three different light sources were used in the experiments, which include Entela multiband UV lamp ($\lambda$=254 or 366 nm), light emitting diode (LED) ($\lambda$=630~680 nm, output <1mW), and florescent light ($\lambda$=400~700 nm). Figure 1(c) shows the chemical structure of the molecules, which were carefully selected to investigate the effects of molecule-length, substituent group and conjugation degree on device performance. The SAMs were made on the Au surface by immersion the substrate in 1 mM molecular solution in pure THF for 24 hrs. For molecule **1** case, the SAMs were grown from **1**: decanethiol solution in THF with mole ratio of 1:1 to reduce the monolayer defect.[18] For each molecule studied, a number of junctions from different samples were characterized at the same conditions. The junctions were stable enough to sustain continuing bias scan (from -1.5 to +1.5 V) for 24 hours at low temperature. Control samples with bare gold junction show linear *I–V* curves that have no response to either light or thermal signals. The current on/off ratio was calculated from the *I–V* curves measured in the dark and under light illumination, respectively.

The charge transport in the molecular crossbar junctions is sensitive to light illumination. Figure 2(a) shows the *I–V* measurements of molecule **1** junction at 95 K. The *I–V* characteristics are strongly asymmetric in the dark, which show an open-circuit behavior at the positive bias range with leakage current of few pA. In contrast, higher conduction and hysteretic loop are observed at the negative bias side. Under 366-nm UV light illumination, the conduction of the junction was significantly increased. The current ratio between light ("1") and dark ("0") states is more than 3 orders of magnitude at + 4 V. The *I–V* curves became more symmetric under light, which can be separated into two characteristic regimes: a linear portion at the lower bias range (from –0.5 to + 0.5 V) and a nonlinear part at the bias beyond. As further discussed, these results can be attributed to the effect of the distinct binding nature of the molecule/electrode interfaces on the charge transport,[10,11,19] although charge trapping/detrapping and Coulomb interactions may also exist in the junction, especially, in the dark and at low temperature conditions.[20]

More interestingly, the molecular junction can be tunned between "0" and "1" states by modulating the light intensity. As demonstrated by figure 2(b), the conduction of molecule **1** junction is stepwise enhanced via increasing the light intensity step-by-step from zero (i.e., dark) to 3.11 cd/m$^2$. The plots of current versus light intensity follow a linear dependence at given bias voltages. The line slope is proportional to the applied voltage, while the y-axis intercept equals the junction current in the dark. Promisingly, such device performance may enable the crossbar junction to serve as a prototype of optically gated molecular transistors. Similar device performances were reproducibly obtained using optical signals with wavelength from 254 to 700 nm. This indicates that the switching should be independent of light wavelength.[21]

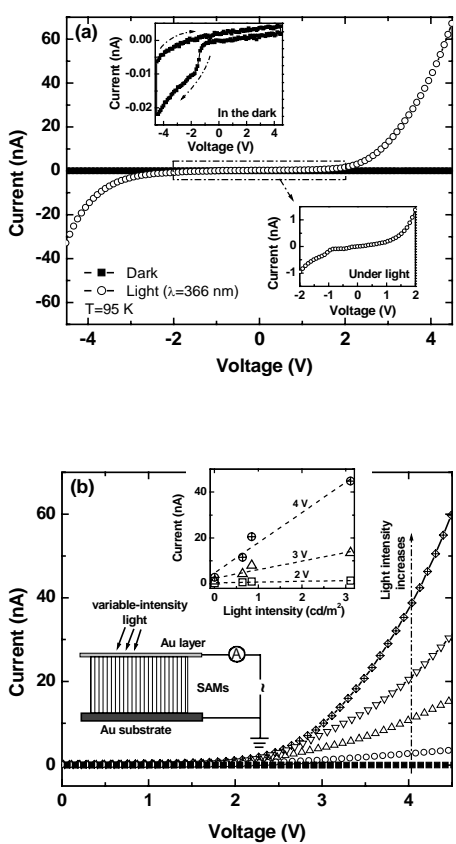

Similar switching was reproducibly obtained in crossbar junctions with very different molecules (see Fig.1(c)), suggesting that the switching has nothing to do with any molecule-specific transition.[22] However, the switching ratio is shown to be dependent on the molecular structure to some extent. The switching ratio of the crossbar junctions follows a general sequence of molecule **1** > dodecanethiol ($C_{12}$) > decanethiol ($C_{10}$) > nonanedithiol ($C_9$) > octanethiol ($C_8$) within ± 2 V. This result is consistent with the observations in STM-tip/SAMs/Au and Hg/SAMs/Ag junctions that conjugate molecules possess a lower decay factor $\beta$ than that of the alkylthiols.[2,3] Because the junction current $I$ can be described with $I=I_0 \exp(-\beta d)$ in low bias voltage. Where, $d$ is molecule-length. The lower the decay factor $\beta$ the higher the junction current $I$.

The switching performance of the crossbar junctions studied is not only reversible but also very stable. Figure 3 shows the current–time curve of nonanedithiol junction at 95 K when the junction was repeatedly exposed to fluorescent light. The junction was stable after many switching cycles. The highest current on/off ratio is about 700 at +1 V. Similar switching is also observed at high temperature. However, the current on/off ratio will drastically decrease to less than 5 as the temperature increases to 270 K, indicating that the switching is strongly dependent on the temperature.

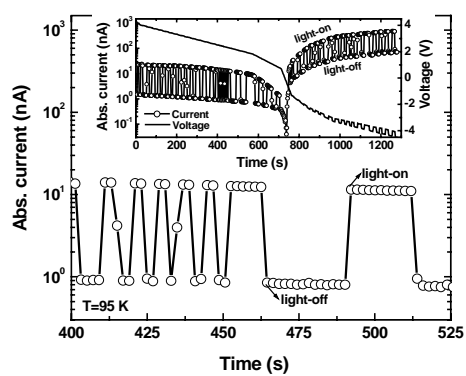

**Figure 2** Molecule 1 junction at 95 K. (a) *I–V* curves measured in the dark and under light illumination, respectively. The lower inset magnifies part of the data under light, while the upper one shows the *I–V* in the dark. Note the dramatic decrease of the current scale in the dark. (b) *I–V* curves plotted as a function of the light intensity. As evidenced by the dashed best linear fit, the plots of current versus light intensity show a linear behavior at given bias voltage (upper inset). The lower inset illustrates a schematic setup of optically gated molecular junction based transistors.

**Figure 3** Reversible switching of nonanedithiol junction at 95 K under the stimulus of florescent light signal. The junction is stable after many switching cycles (inset).

To understand the switching mechanism, the temperature dependence of the *I–V* characteristics was investigated for all the molecules shown in Fig. 1 (c). It is found that the junction performance is strongly dependent on the temperature. Figure 4 plots the *I–V* curves of decanethiol junction as a function of the temperature. The current ratio between 91 K and 300 K is about $10^3$ at + 4V in the dark. The ratio drastically decreased to about 50 under light illumination, suggesting that the light may induce a remarkable change in the charge transport. Furthermore, the *I–V* temperature dependence is shown to be dependent on the bias direction under light condition. They are more dependent on the temperature at the positive voltage than that at the negative side. This observation is consistent with the fact that the electrode/molecule interfaces of the crossbar junctions have distinct chemical nature and thus have extraordinarily different electron tunneling parameters. Initial *I–V* data analysis indicates that the charge transport in the junction can be mainly attributed to thermionic emission process under light condition,[13] while Fowler-Nordheim tunneling might play a dominant role in the dark. It confirms that Fowler-Nordheim tunneling can be a temperature-dependent, although the mechanism may be different.[23]

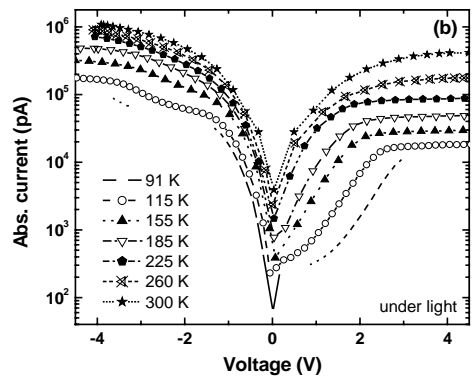

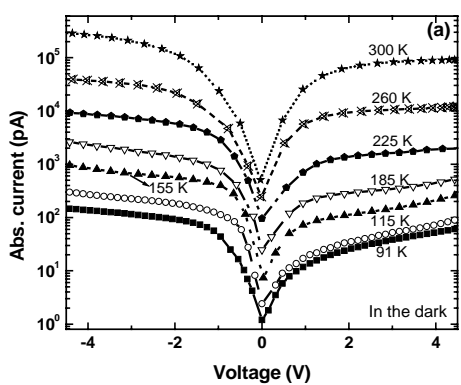

**Figure 4** *I–V* temperature dependence of decanethiol junction (a) in the dark and (b) under illumination of florescent light.

One plausible explanation for the switching is that the incident light may induce a change(s) in the electron tunneling parameters at the top molecule/electrode interface. Several facts support this hypothesis: (1) No molecular structure dependence was observed for the switching, confirming that it has not resulted from any molecular transitions. (2) As a clear indication, the temperature dependence of the junction *I–V* was very different between dark and light conditions. (3) No similar performances were observed in bare gold junctions and thus the possibility of artificial and substrate effects can be excluded. Another possible interpretation is the change of the 'effective' junction size and/or other local environment. Since the top electrodes of the crossbar junctions were transferred onto the substrate in solid-state from the PDMS stamp, there should exist an abrupt difference between the bottom (strong S-Au bond) and the top (weak physisorption) molecule/electrode interfaces. As a result, only a fraction of the molecules in the junction area can consequently form a 'firm' contact with the top PDMS stamped electrode, even though all of them were chemically bonded with the bottom one. Under such situation, the effective junction size and/or the electron tunneling rate at the top interface may be greatly increased when the junction is exposed to the stimulus of either thermal or light signals.[24] This simple 'picture' shows that the electrode/molecule interfaces may be very different for the same molecular junction fabricated using different approaches. Also, it explains very well why the current density of the crossbar junctions is much lower than that of the junctions fabricated using other methods.[2-6,24] The top interface may act as a bottleneck for charge transport,[25,26] which can consequently lead to electrical rectification in the junctions.[27-30]

In summary, reversible optoelectronic switching is observed in addressable molecular crossbar junctions fabricated using PDMS stamp-printing method. The switching is independent of both molecular structure and optical wavelength, while it strongly depends on the temperature. It is shown that the distinct binding nature of the molecule/electrode interfaces plays a dominant role in the electrical performance of the crossbar junctions. The stamp-printing approach may provide a potential way to develop novel molecular junction based devices like switches, transistors, and light sensors.

The author thanks the support and help from Luping Yu, Heinrich Jaeger, Shengwen Yuan, G. M. Morales, and Arturo Sanchez.